# Modulation of bandgap in bilayer armchair graphene ribbons by tuning vertical and transverse electric fields

Thanh-Tra Vu[1], Thi-Kim-Quyen Nguyen[2], Anh-Huy Huynh[1], Thi-Kim-Loan Phan[1], Van-Truong Tran[3, 4*]

[1]Department of Physics, School of Education, Can Tho University, Can Tho, Vietnam

[2]School of Graduate, College of Natural Sciences, Can Tho University, Can Tho, Vietnam.

[3]C2N, Université Paris-sud, Université Paris Saclay, CNRS, 91405 Orsay, France

[4]EM2C, CentraleSupélec, Université Paris Saclay, CNRS, 92295 Châtenay Malabry, France

*van-truong.tran@u-psud.fr

We investigate the effects of external electric fields on the electronic properties of bilayer armchair graphene nano-ribbons. Using atomistic simulations with Tight Binding calculations and the Non-equilibrium Green's function formalism, we demonstrate that (i) in semi-metallic structures, vertical fields impact more effectively than transverse fields in terms of opening larger bandgap, showing a contrary phenomenon compared to that demonstrated in previous studies in bilayer zigzag graphene nano-ribbons; (ii) in some semiconducting structures, if transverse fields just show usual effects as in single layer armchair graphene nano-ribbons where the bandgap is suppressed when varying the applied potential, vertical fields exhibit an anomalous phenomenon that the bandgap can be enlarged, i.e., for a structure of width of 16 dimer lines, the bandgap increases from 0.255 eV to the maximum value of 0.40 eV when a vertical bias equates 0.96 V applied. Although the combined effect of two fields does not enlarge the bandgap as found in bilayer zigzag graphene nano-ribbons, it shows that the mutual effect can be useful to reduce faster the bandgap in semiconducting bilayer armchair graphene nano-ribbons. These results are important to fully understand the effects of electric fields on bilayer graphene nano-ribbons (AB stacking) and also suggest appropriate uses of electric gates with different edge orientations.

.

# 1. Introduction

Graphene, a fantastic 2D material has been demonstrated as a promising material for various applications in different fields, particularly in electronics due to its spectacular and outstanding electronic properties with an extremely high electron mobility thanks to the linear behavior of the energy band structure near the neutrality point.[1–3] In spite of huge potential applications, graphene suffers from a lack of a bandgap.[2,4,5] Since the energy bandgap plays a crucial role in the operation of semiconductor devices such as p–n junctions, transistors, and sensors,[6–8] a tunable bandgap is thus highly desirable to flexibly optimize performance of such devices.

To exploit the potentials of graphene for these applications, a numerous studies have been therefore carried out to overcome this substantial problem. Among many strategies that have been proposed such as engineering graphene by doping some outer atoms,[9,10] introducing nano-holes,[5,11] or creating hybrid structures of graphene and other similar materials,[12–14] the use of external fields is still one of the most effective methods to open and control the bandgap of graphene since electric gates are easy to be set up and the effective strength is readily controlled by varying the bias between gates.[15,16]

In 2006, McCann and Fal'ko[17] effectively applied this technique and theoretically demonstrated that a tunable bandgap can be open in bilayer structures of 2D graphene sheets in the presence of a vertical electric field. This exciting prediction was confirmed experimentally later by Ohta et. al.[18] Further studies have shown that a largest bandgap about 0.25 eV can be achieved in 2D bilayer graphene with the technique of applying vertical electric fields.[19–22]

Although transverse electric fields are not appropriate for 2D structures, it has been revealed that this type of electric fields is very relevant in modulating the electronic properties of ribbon form of graphene.[23–26] In ref. [23], Chang et. al. showed that the bandgap of single layer graphene nano-ribbons (SL-GNRs) can be tuned remarkably by transverse electric fields, i.e., the bandgap is open in the case of semi-metallic single layer armchair and zigzag ribbons (SL-AGNRs and SL-ZGNRs) [23] but it is reduced strongly in the case of semiconducting ribbons.

Recently, a combination of a vertical and a transverse electric fields in bilayer graphene nano-ribbons (BL-GNRs) has been presented with interesting phenomena.[27] It has shown that the bandgap in bilayer zigzag graphene nano-ribbons (BL-ZGNRs) is largest under the simultaneous effect of both fields. However, the impact of these external fields on bilayer armchair structures (BL-AGNRs) are still pending.

Unlike the case of zigzag structures in which the bandgap is equal to zero regardless of ribbon width,[28–30] it has found that the electronic properties of structures with armchair edges, more precisely SL-AGNRs, strongly depend on width, i.e., the bandgap is classified into three families $3p$, $3p + 1$, $3p + 2$ with $p$ as an integer number.[28,31] It is thus physics underlying the effect of external electric fields on the electronic properties of BL-AGNRs must be richer compared to that of BL-ZGNRs and need to be unveiled.

In this article, we investigate the individual and combined impacts of a vertical and a transverse electric fields on the electronic properties of BL-AGNRs. By using atomistic calculations with a tight-binding (TB) model and the non-equilibrium Green's function (NEGF) formalism, we show that bandgap is open in the case of semi-metallic BL-AGNRs (group $3p + 2$) and suppressed in almost all cases of other family structures with either field applied. Such phenomena are similar to those obtained in SL-GNRs and also in BL-ZGNRs. However, the outcome from group $3p + 1$ is an exception since bandgap can even be enlarged (not reduced as in other semiconducting structures) under the effect of vertical electric fields. More interestingly, whereas transverse electric fields have been demonstrated to be more effective than vertical ones in BL-ZGNRs in terms of inducing larger bandgap,[27] here we show an inverse phenomenon, i.e., larger bandgap with vertical fields.

## 2. Modeling and Methodologies

### 2.1. Modeling

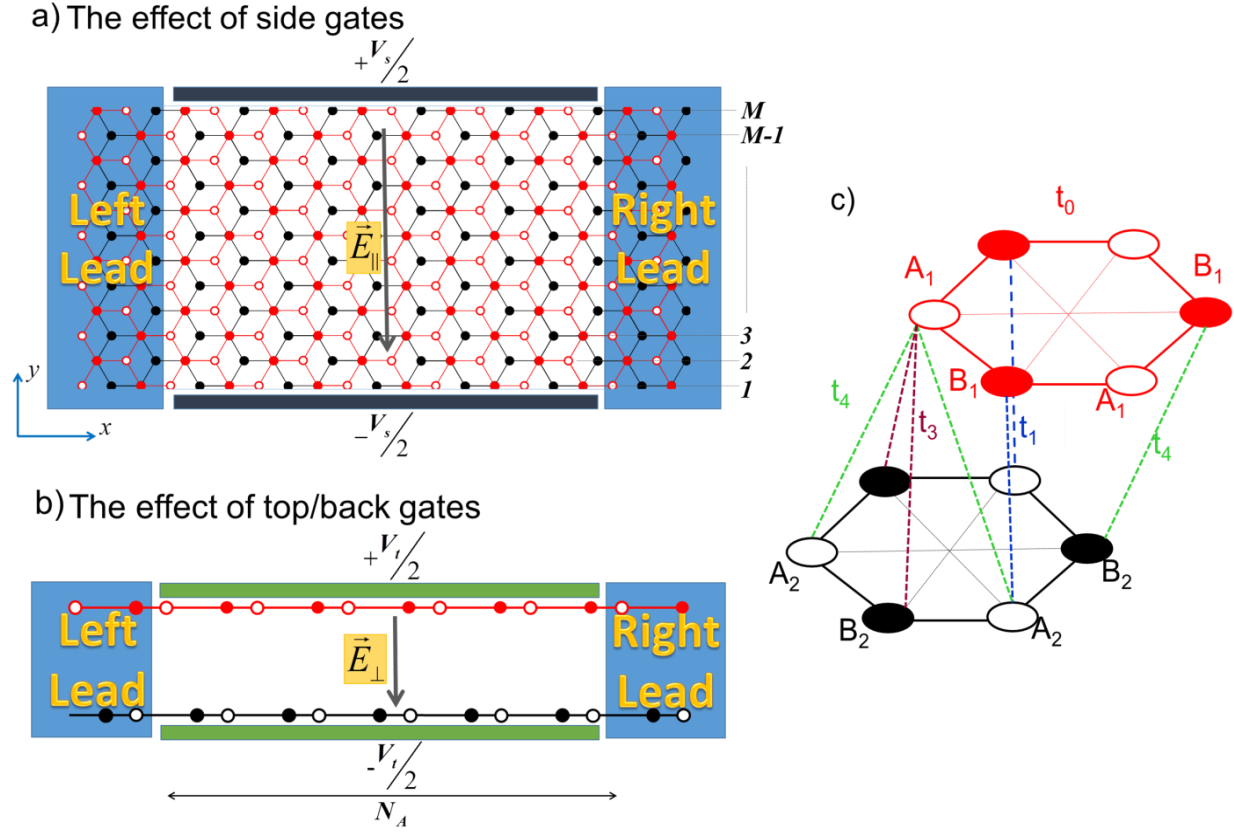

**Figure 1.** Schematic views of structures made of BL-AGNRs (AB stacking): (a) a BL-AGNR under the effect of a transverse electric field generated by side gates $+V_s/2$ and $-V_s/2$. (b) The structure placed in the region of a vertical electric field generated by a top gate $+V_t/2$ and a bottom gate $-V_t/2$. (c) The diagram for intra-layer and inter-layer interactions between atoms in a single layer and in two layers.

In this present work, we consider BL-AGNRs in the presence of a transverse or/and a vertical electric field as schematized in Fig. 1(a) and Fig. 1(b), respectively. We assumed that a vertical electric field $\vec{E}_\perp$ can be generated by top $+V_t/2$ and back $-V_t/2$ gates, while a transverse electric field $\vec{E}_\parallel$ can be induced by two side gates with potentials $+V_s/2$ and $-V_s/2$. The width of BL-AGNRs is characterized by number of dimer lines $M$ along the width of each sub-ribbon. In Fig. 1(c) we sketched the typical stacking between the two layers (AB or Bernal stacking) where $B_1$ sites of the upper layer (red) are located exactly on the top of $A_2$ sites of the lower layer (black), and $A_1$ or $B_2$ sites are above or below the center of hexagons in the other layer. Also in this figure, interactions between atoms are also illustrated by a set of hoping parameters $\{t_0, t_1, t_3, t_4\}$ in which $t_0$ is the intra-layer hoping energy between two nearest atoms in a single layer, while $t_1$, $t_3$ and $t_4$ are the inter-layer hoping energies describe interactions between atoms in different layers.[32–34]

### 2.2. Methodologies

To investigate the electronic properties of BL-AGNRs, a TB model was employed and the Hamiltonian is generally written as

$$H = \sum_i U_i |i\rangle\langle i| - \sum_{\langle i,j \rangle} t_{ij} |i\rangle\langle j|, \quad (1)$$

where $U_i$ is the total energy potential at $i$-th site and determined by expression $U_i = \mp e.V_t/2$ in the case of a vertical field applied. The signs "-" and "+" are for atoms belong to the upper and lower layers, respectively. And $U_i = -e.(-V_s/2 + E_\parallel . y_i)$ in the case of a sole transverse field applied, where $E_\parallel = V_s/W$ is the average strength of the transverse field and $y_i$ is the coordinate of atom $i$-th with respect to the side gate – $V_s/2$. Also in equation (1), $t_{ij}$ is the coupling between atom at $i$-th site and its surrounded neighbor atoms and it will be fit to $t_0$, $t_1$, $t_3$ or $t_4$ depending on the level of the distance between the two atoms. In our calculations, the hoping energies were taken from ref. [32] where they were parameterized to fit *ab* initio results, i.e., $t_0 = 2.598$ eV, $t_1 = 0.364$ eV, $t_3 = 0.319$ eV and $t_4 = 0.177$ eV.

Transport properties of the structures were also investigated by the NEGF approach. Within this formalism, the transmission coefficient $T(E)$ is calculated by equation[35]

$$T = \text{Trace}\left(\Gamma_L G_D \Gamma_R G_D^\dagger\right), \quad (2)$$

where $G_D = [E - H_D - \Sigma_L - \Sigma_R]^{-1}$ is the Green's function of the active region (device), and $\Gamma_{L(R)} = i\left(\Sigma_{L(R)} - \Sigma^\dagger_{L(R)}\right)$ is considered as the injection rate at the interfaces of the left (right) leads and the active region.[35,36]

Besides, to support for the energy band analysis, we have also calculated the local and total density of states (LDOS and TDOS), which are defined by[37]

$$LDOS_i(E) = -trace\left[\frac{Im(G_{ii})}{\pi}\right] \quad (3)$$

and

$$TDOS(E) = -trace\left[\frac{Im(G)}{\pi}\right], \quad (4)$$

where $G$ is the Green's function of the Hamiltonian used to calculate the band structure.

## 3. Results and discussions

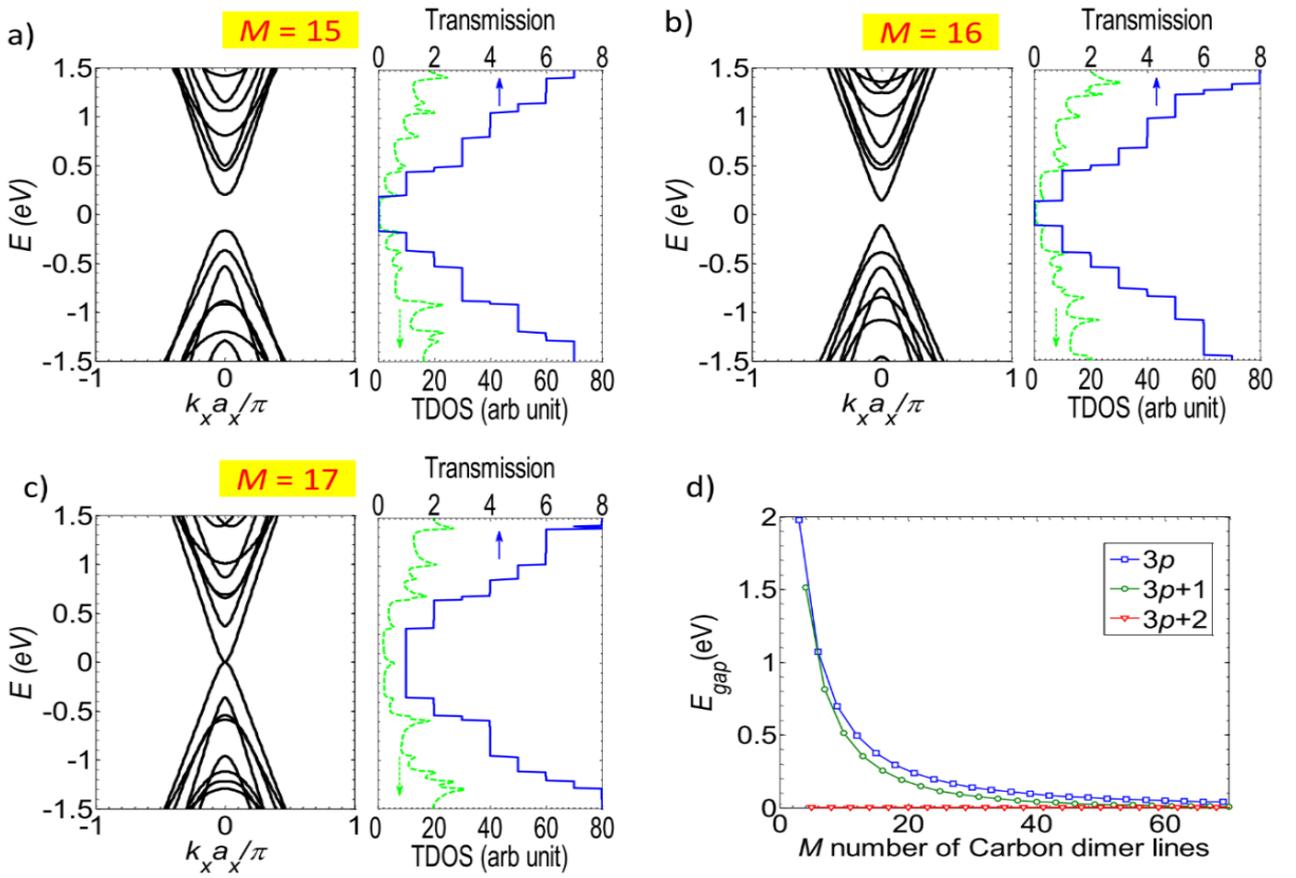

**Figure 2:** The band structure and the TDOSs of BL-AGNRs with different widths (a) $M = 15$, (b) $M = 16$, (c) $M = 17$. (d) The bandgap is plotted as a function of number of dimer lines $M$ along the ribbon width for three groups $3p$, $3p + 1$, and $3p + 2$.

In this section we will analyze the electronic properties of BL-AGNRs, first in the absence and then in the presence of different external electric fields. Since the bandgap and other electrical quantities are dependent on different families of ribbon width, unless otherwise stated, we focus the investigation on three ribbons of width $M = 15, 16\ 17$ which characterize three groups $M = 3p, 3p + 1, 3p + 2$, respectively. Moreover, for analysis of energy bands, the active region with the effects of external fields was consider to be long enough so that this path can behave as a periodic structure.

### 3.1. Band structure analysis of BL-AGNRs

We first examine the electronic properties of pristine BL-AGNRs (without fields applied). The calculated band structure, the TDOS and the transmission coefficients for all three ribbons are shown in Fig. 2. It can be seen from energy bands in Fig. 2(a) and Fig. 2(b) that the bandgaps are presented in the structures of width $M = 15$ and $M = 16$. This result is also reflected clearly in the spectrum of the TDOSs (dashed green lines) or of the transmissions (solid blue lines) where zero density of states or zero transmission can be observed around the energy point $E = 0$. It is worth to note that the peaks of the TDOSs or the transitional positions of the transmissions happen at the bottom (top) of conduction (valance) bands. In a comparison, the bandgap of the structure of width $M = 15$ seems larger than that of the structure with $M = 16$. In fact, a bandgap about 0.374 eV was found in the smaller bilayer ribbon and about 0.255 eV in the bigger one. In contrast, this value is almost equal to zero for the case $M = 17$ as seen in Fig. 2(c), and in this case the TDOS shows a peak at the touching point (Dirac point) of the lowest conduction and highest valance bands, while the transmission coefficient always presents a finite-value due to semi-metallic properties of the ribbon. Thus the bandgap of BL-AGNRs strongly depends on the width of sub-ribbons as in SL-AGNRs.[38]

To further explore the width dependence of the bandgap and characteristics of different family structures $M = 3p, 3p + 1, 3p + 2$, we have simulated other ribbons and plotted the bandgap as a function of ribbon width. The results are displayed in Fig. 2(d). It can be observed clearly that, for group $M = 3p + 2$ the bandgap is almost equal to zero, whereas the bandgap is opened for ribbons belong to groups $M = 3p + 1$ and $3p$. Moreover, the bandgap corresponding to group $3p$ is slightly larger than that of group $3p + 1$. In the range of the number of dimer lines $M$ from 3 to 15, the bandgap of both groups $3p$ and $3p + 1$ drops remarkably but the value weakly reduces in longer structures, for example, for group $3p$, $E_{gap}$ decreases about 1.603 eV from 1.977 eV in the bilayer ribbon of width $M = 3$ to 0.374 eV in the case $M = 15$, but the difference is just about 0.209 eV if we go from $M = 15$ to $M = 27$. For sufficient long ribbons, the bandgap of these two groups tends to approach zero and eventually becomes semi-metallic as for group $3p + 2$. This result is in agreement with what has been observed in the case of 2D bilayer graphene structure.[39]

### 3.2. The distortion of the band structure under the effects of external electric fields

In this section, the impacts of vertical and transverse electric fields on energy bands are considered.

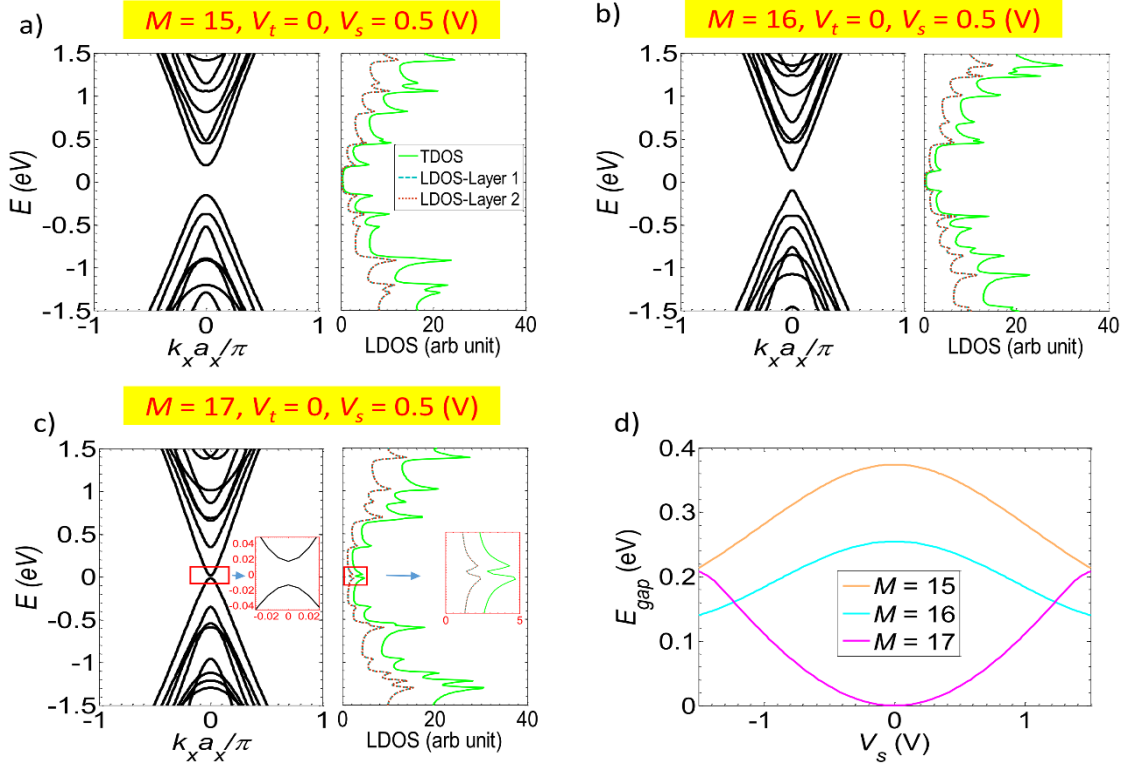

**Figure 3.** The effect of transverse electric fields on energy bands and the LDOSs, the TDOSs of the three considered structures (a) $M = 15$, (b) $M = 16$ and (c) $M = 17$. (d) The bandgap is considered as a function of the side gate potential $V_s$.

#### 3.2.1. The effect of transverse electric fields

In Fig. 3, we display energy bands and the TDOSs of the three structures under the effect of a transverse electric field with a potential strength $V_s = 0.5$ V. To determine the field effect on each sub-ribbon layer, we also plot the LDOSs of each layer. The calculated bandgaps of the two structures $M = 15, 16$ are $E_{gap} = 0.346$ eV and $0.233$ eV, respectively which are thus smaller than those obtained in the case without fields applied. This reduction of the bandgap is also reflected from the narrowing of the valleys of the TDOSs as observed in Fig. 3(a) and Fig. 3(b). However, in Fig. 3(c), for the structure of width $M = 17$, the bandgap tends to open, which is indicated by a small gap about 30 meV appearing in the band structure. This slightly change of the bandgap can be observed more clearly from the spectrum of the density of states where we can see two separated peaks near zero energy instead of one as we observed previously in Fig. 2(c).

To exploit the potential to tune the energy gap in these structures, we investigated in more details the dependence of the bandgap opening on the applied voltage $V_s$ and show it in Fig. 3(d). As it can be observed, the variations of bandgap are symmetrical for negative and positive values of $V_s$ and look like parabolic shapes. In the considered range of $V_s$ from 0 to 1.5 V, the bandgap is suppressed in the structures $M = 15, 16$ but it is open and enlarged in the case $M = 17$, i.e., the bandgap is dropped from 0.374 eV to 0.216 eV in the case of $M = 15$ and from 0.255 eV to 0.140 eV in the case of $M = 16$, while it increases from almost zero to 0.208 eV in the structure $M = 17$. Substantially, these phenomena are similar to those obtained in the case of SL-AGNRs under the effect of transverse electric field.[23]

It is also worth to note that the LDOSs indicate that the effects of the transverse field on two layers are the same everywhere. This behavior can be understood owing to the relative positions of the two layers are the same with respect to the side gates as shown in Fig. 1(c).

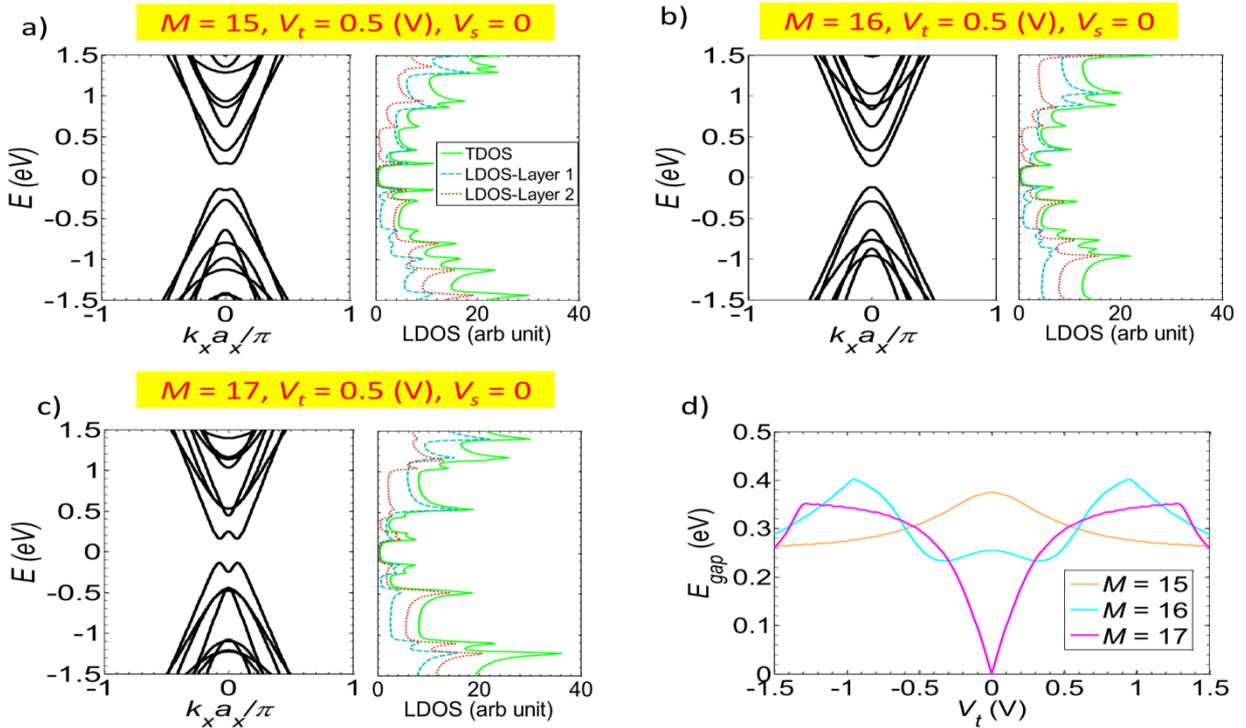

**Figure 4.** The effect of vertical electric fields on energy bands and the LDOSs and the TDOSs in the three considered structures (a) $M = 15$, (b) $M = 16$ and (c) $M = 17$. (d) Evolution of the bandgap in the three structure as varying the top/back gate potential $V_t$.

### 3.2.2. The effect of vertical electric fields

To reveal the difference of a vertical field from a transverse one, in this sub-section we re-examine the change of the band structure as considered in Fig. 4 but now under the influence of a vertical field. In Fig. 4, we show the band structures, the DOSs and the LDOSs of the considered structures when a top/back potential $V_t = 0.5$ V is applied. First, from the panels including the LDOSs we can see that the LDOSs are no longer superimposed as in the case of transverse fields, but they are totally separated and shifted in opposite directions due to the opposite effects of the vertical field on the two layers. This phenomenon is similar to that observed in the case of BL-ZGNRs.[27]

Regarding to the features of the bandgap, we obtained a gap about 0.309 eV in the structure $M = 15$ which is thus smaller than that of the pristine structure. Surprisingly, the structure $M = 16$ exhibits a bandgap about 0.261 eV, which is slightly larger than that of the counterpart without field applied (0.255 eV) and actually it is contrary to what obtained above for the case of transverse fields. Moreover, the band structure of the bilayer ribbon $M = 17$ in Fig. 4(c) presents a bandgap of 0.297 eV which is much larger than that obtained in Fig. 3(c). This latter result predicts that in the case of BL-AGNRs, vertical electric fields could be more effective than transverse ones in terms of opening larger bandgap.

To fully understand the dependence of the bandgap on the strength of the vertical field, we examined and plotted the bandgap as a function of $V_t$ as shown in Fig. 4(d). Substantially, the variation in the structure $M = 15$ is similar to that in the case under the effect of transverse fields. However, the behavior is unusual in the case of $M = 16$ as we observe a fluctuation of the bandgap over the potential range, i.e., first, bandgap reduces from 0.255 eV at $V_t = 0$ to 0.234 eV at the absolute value of potentials $/V_t/ = 0.36$ V, then it turns and increases up to a maximum value $(E_{gap})_{max} = 0.40$ eV when the applied potential reaches $/V_t/ = 0.96$ V before falling down. This is an exciting result as bandgap of a semiconducting structure can be enlarged by vertical electric fields and thus it is different from the monotonic reduction of the bandgap observed in the case of transverse electric fields in Fig. 3(d) as well as in the case of SL-AGNRs.[23,26] For the structure of width $M = 17$, the bandgap increases significantly in the range of [0 V, 0.5 V], then the growth is slower and the bandgap reaches a peak of $(E_{gap})_{max} = 0.35$ eV at $/V_t/ = 1.3$ V. This value is obviously larger than 0.208 eV, the one we obtained under the impact of transverse fields and confirms that vertical fields are more effective than transverse ones in BL-AGNRs as predicted above.

It is also noted that the factual behavior of curves in Fig. 4(d) is actually associated directly with the distortion at the bottom and top of the lowest conduction and highest valance bands, respectively, as can be observed more clearly in Fig. 4(a) and Fig. 4(c). These distortions lead to the fact that the bangap is no longer direct at the Gamma point, but somewhere else between the Gamma point and the boundaries of the 1st Brillouin zone.

### 3.2.3. Comparisons between the two electric fields

It has been demonstrated that in the case of BL-ZGNRs, the combined effect of both transverse and vertical electric fields can open larger bandgap compared to a sole field applied.[27] This result motivates a further investigation of the effect of the two fields in the case of BL-AGNRs considered here.

To have a fair comparison with phenomenon happening in BL-ZGNRs, we first examine the mutual effect on the semi-metallic structure $M = 17$ and the bandgap is now plotted as a function of both $V_s$ and $V_t$. The result is displayed in Fig. 5(a). From the color distribution in the figure, it can be seen clearly that the largest bandgap is found in the axis $V_s = 0$, that is not in the regions where two fields are applied simultaneously. It indicates that in the case of semi-metallic BL-AGNRs, the combined effect does not induce larger bandgap than in the case a single vertical field is applied.

A similar conclusion can be seen in the case of $M = 16$ in Fig. 5(b) where the maximum bandgap is just obtained at $V_t = \pm 0.96$ V and $V_s = 0$. Although the combination of two fields does not enlarge bandgap in this case, it is still useful if we want to reduce bandgap as we see dark-blue color near the corners of the figure.

There is no enhancements of bandgap in the structure $M = 15$ as we can see in Fig. 5(c). In terms of effectiveness in reduction of the bandgap, it seems that transverse fields are a bit better than the vertical fields as the bandgap decreases faster along $V_s$ axis ($V_t = 0$). Similar to the case of $M = 16$, the regions near the corners also indicate that the simultaneous impact of the two fields can reduce bandgap more rapidly compared to a single field one.

In addition, Fig. 5(a) exhibits that vertical fields induce larger bandgap than transverse fields. And Fig. 5(b) clearly presents that vertical fields can enlarge the bandgap of this semiconducting structures, while transverse fields only reduce the bandgap. These points are consistent with results in Fig. 3 and Fig. 4.

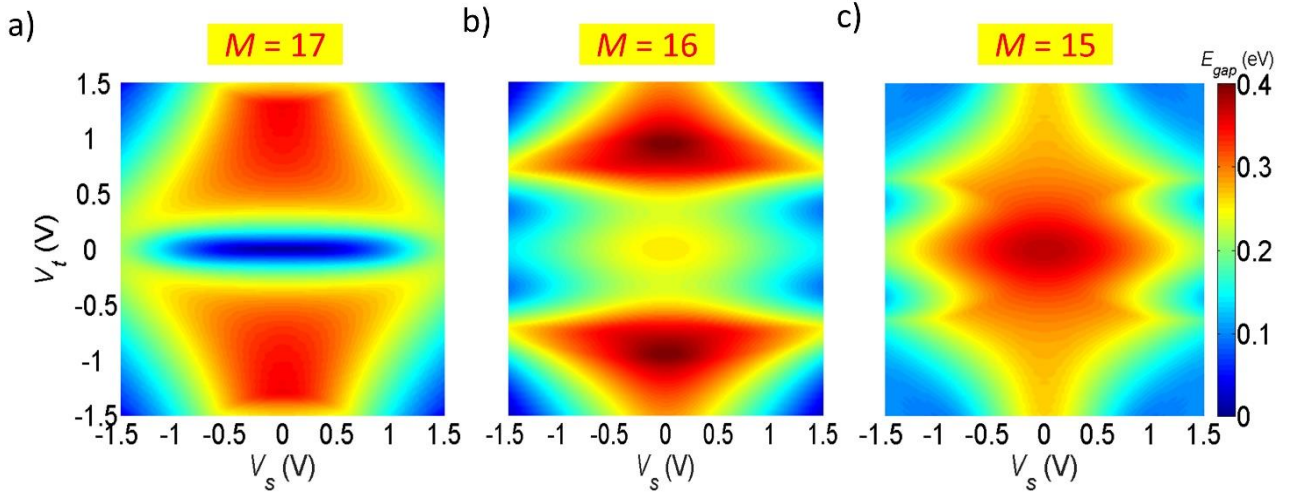

**Figure 5.** The bandgap spectrum of different BL-AGNRs as varying both vertical and transverse fields. Results obtained for (a) $M = 17$, (b) $M = 16$ and (c) $M = 15$.

## 4. Conclusions

In the present paper, we have studied the electronic properties of BL-AGNRs both without and with external electric fields applied. In the absence of external fields, it has shown that the bandgap of BL-AGNRs is apparently classified in to three groups $3p$, $3p + 1$, $3p + 2$ in which group $3p + 2$ is semi-metallic while the others are semiconducting as in the case of SL-AGNRs. In the presence of external fields, first it has demonstrated that the impact of transverse fields on the band structure of BL-AGNRs is similar to that of SL-AGNRs. However, in the case of vertical fields applied, two interesting outcomes have been pointed out: (i) vertical fields are more effective than transverse fields in terms of opening larger bandgap in semi-metallic structures, which is thus contrary to the effect demonstrated in BL-ZGNRs in previous studies; (ii) for semiconducting group $3p + 1$, vertical fields can enlarge the bandgap, i.e., for $M = 16$, $E_{gap}$ increases from 0.255 eV to the maximum value of 0.40 eV. Meanwhile transverse fields only induce a reduction of the bandgap similar to that obtained in SL-AGNRs.

By considering the combined effect of the two fields we have also shown that the simultaneous use of the two fields is not better than a single vertical field in terms of opening or enlarging the bandgap. However, the mutual effect can be useful to reduce faster the bandgap in semiconducting BL-AGNRs.

Our results are important to fully understand the effects of electric fields on BL-GNRs (AB stacking) and also suggest appropriate uses of electric gates with different edge orientations of these structures.

**Acknowledgments**

This research is funded by Vietnam National Foundation for Science and Technology Development (NAFOSTED) under grant No. 103.01-2015.98. We also would like to acknowledge the funding support from Natural Science Foundation of Cantho University under grant No. 1012/QĐ-ĐHCT.